\documentclass[preprint,aps]{revtex4}
\newcommand{\be}{\begin{equation}}
\newcommand{\ee}{\end{equation}}
\newcommand{\ba}{\begin{eqnarray}}
\newcommand{\ea}{\end{eqnarray}}
\newcommand{\p}{\partial}
\newcommand{\f}{\frac}

\begin{document}
\title
{On adiabatic evolution for a general time-dependent quantum system}

\author{Kyu Hwang Yeon$^1$, Jeong Ryeol Choi$^2$, Shou Zhang$^3$, and Thomas F. George$^4$
 \vspace{0.5cm}}

\address
{$^1$BK21 Physics Program and Department of Physics,
College of Natural Science,  Chungbuk National University,
 Cheongju,  Chungbuk 361-763, Republic of Korea \vspace{0.2cm}}

\affiliation{$^2$Department of Radiologic Technology,~Daegu Health College, Taejeon 1-dong, Daegu 702-722, Republic of Korea \vspace{0.2cm}}

\address
{$^3$ Department of Physics, College of Science, Yanbian University, Yanji, Jilin 133002,
People¡¯s Republic of China
\vspace{0.2cm}}

\address
{$^4$
Office of the Chancellor and Center for Nanoscience, Department of Chemistry \& Biochemistry and Department of Physics \& Astronomy, University of Missouri-St. Louis, St. Louis, Missouri 63121, USA
\vspace{0.2cm}}

\begin{abstract}

The unitary operator corresponding to the classical canonical transformation that connects a general
closed system to an open system under adiabatic conditions is found. The quantum invariant
operator of the adiabatic open system is derived from the unitary transformation of the quantum
Hamiltonian of the closed system. On the basis of these results, we investigate the evolution of
the general quantum adiabatic system and construct a revised adiabatic theorem.
The adiabatic theorem developed here exactly reduces to the well-known Berry adiabatic theorem when
the control parameter
of an adiabatic system is constant in time. \\
\\
\end{abstract}
\maketitle \vspace{0.0cm}

\section{Introduction}
Although quantum states of a closed conserved system are given as solutions of the Schr\"{o}dinger
equation, there is no general method to derive quantum states of an open quantum system that is
affected by an external environment\cite{Bellac, Sak}.
However, an open quantum system that deviates slightly from a closed system may
satisfy the adiabatic condition and, consequently, we can regard the Schr\"{o}dinger solutions as
the corresponding quantum states of the system.
A well-known method to treat this problem is Berry's adiabatic theorem, which states
that ``the quantum states of an adiabatic open quantum system are proportional to the eigenstates of
its Hamiltonian"\cite{Sak,Ber, Alf}.
Active searches for Berry's phase have been undertaken \cite{Biao,Kra,Gon} on account of its
importance in various fields of theoretical physics such as Aharonov Bohm oscillations \cite{jeng},
quantum dots \cite{vag}, quantum Hall effect \cite{kyu} and geometric phase gate \cite{ajm}.

Here, we obtained the unitary operator that connects quantum states of an open quantum system
that depends on time-varying external parameters with those of a conserved system. We demonstrate
that the quantum states of an adiabatic open quantum system can be
derived from those of a conserved system.

Though the invariant operator of an adiabatic open quantum system can be found from a straightforward evaluation,
 it is also possible to evaluate this operator via a unitary transformation of the Hamiltonian
of a conserved system.
From this we want to demonstrate that the
quantum states of an adiabatic open quantum system are proportional
to the eigenstates of the invariant operator of the system.
Among various types of invariant operators for a quantum system, the quadratic invariant operator
is the same as the Hamiltonian when the control parameter does not vary with time.
Thus, we can view this concept as an extended version of Berry's adiabatic theorem. 

\section{A quantum system with time-dependent external parameter}
   We first  consider a linear classical transformation given as
\be
 \left\{\begin{array}{ll}
q(t) = e^{\alpha(t)}Q(t) - \beta(t), \\
p(t) = e^{-\alpha(t)}P(t) + m \dot{\alpha}(t) e^{\alpha(t)}Q(t), \label{1}
\end{array}\right.
\ee
or
\be
 \left\{\begin{array}{ll}
Q(t) = e^{-\alpha(t)}[q(t)+\beta(t)] , \\
P(t) = e^{\alpha(t)}p(t) - m \dot{\alpha}(t) e^{\alpha(t)}[q(t)+\beta(t)], \label{2}
\end{array}\right.
\ee
where $\alpha(t)$ and $\beta(t)$ are real functions of $t$ and are differentiable with respect to $t$.
From now on, for simplicity, we do not explicitly display the time-variable dependence for the time
functions $\alpha$, $\beta$, etc. except for some special cases.
 In Eqs. (\ref{1}) and (\ref{2}),
$q$ and $p$ are canonical variables of the system whose Hamiltonian is given in the form
\be
H_1 (q,p) = \f{p^2}{2m} + V(q, R_0),    \label{3}
\ee
where $R_0$ is an external parameter at  $t = 0$. If the transformation of Eqs. (\ref{1}) and
(\ref{2}) is canonical,
$Q$ and $P$ are canonical variables of the system described by the following Hamiltonian \cite{Sud,yeon98}:
\ba
H_2 (Q,P,t) &=& \f{e^{-2\alpha}}{2m}P^2 + \f{m}{2} (\dot{\alpha}^2 + \ddot{\alpha})e^{2\alpha}Q^2
+\dot{\beta}e^{-\alpha}P + m \dot{\alpha}\dot{\beta}e^{\alpha}Q \nonumber \\
& &+ V(e^{\alpha}Q-\beta, R_0).   \label{4}
\ea
Let us assume, in the spirit of a canonical transformation, that the system of the Hamiltonian
of Eq. (\ref{4}) is physically changed from that of Eq. (\ref{3}) by environmental influences
of an external driving force and/or  dissipative frictional force.
Here, we consider only the case of $\alpha(0)=0$,
$\beta(0) = 0$, $\alpha(t) \approx 0$,
$\beta(t) \approx 0$  and $\dot{\alpha}, \dot{\beta} \ll \dot{q}$, so that the change of the
system associated with
the canonical transformation of Eq. (\ref{1}) [or Eq. (\ref{2})] is  adiabatic.
If this condition is satisfied, Eq. (\ref{1})
[or Eq.  (\ref{2})] is the classical adiabatic relation.

Now, we would like to find  the unitary transformation corresponding to the canonical
transformation of Eq. (\ref{1}) [or Eq. (\ref{2})].
The quantum Hamiltonians  $\hat{H}_1$ and $\hat{H}_2$ corresponding to the classical Hamiltonians $H_1$ and $H_2$
can be obtained through the replacement of canonical variables by their corresponding quantum operators
from Eqs.(3) and (4),  respectively. Since the $\hat{H}_1$ system is closed, its quantum state,
$|\psi \rangle$,  can be obtained from the Schr\"{o}dinger equation,
\be
i\hbar \f{\p}{\p t} |\psi \rangle =  \hat{H}_1 |\psi \rangle .  \label{5}
\ee
Since the $\hat{H}_2$ system depends on time, there is no general method to derive the exact quantum states of the system.
However, under the adiabatic condition, we can assert
that the quantum state of the system, $|\Psi \rangle$, obeys the Schr\"{o}dinger equation,
\be
i\hbar \f{\p}{\p t} |\Psi \rangle =  \hat{H}_2 |\Psi \rangle .  \label{6}
\ee

Let us find  an unitary operator $\hat{U}$ that connects the two quantum states associated with the $\hat{H}_1$
and $\hat{H}_2$ systems \cite{yeon98, yeon02,yeon03}:
\be
|\Psi \rangle = \hat{U} |\psi \rangle .  \label{7}
\ee
From Eq. (7), we obtain a differential equation for the operator from Eqs. (5) and (6) as
\ba
i\hbar \f{\p}{\p t} \hat{U} &=& \hat{H}_2 \hat{U} - \hat{U}\hat{H}_1 \nonumber \\
&=&\left( -\f{\dot{\alpha}}{2}(\hat{q}\hat{p} +\hat{p}\hat{q}) + \dot{\beta}e^{-\alpha} \hat{p}
+ m \dot{\alpha}\dot{\beta}e^{\alpha}\hat{q} + \f{m}{2} \ddot{\alpha}e^{2\alpha} \hat{q}^2\right)
\hat{U} .  \label{8}
\ea
We assume that $\hat{U}$ gives the following relations for $\hat{q}$ and $\hat{p}$:
\be
\left\{\begin{array}{ll}
\hat{U} \hat{q} \hat{U}^\dagger = e^{\alpha}\hat{q} - \beta, \\
\hat{U} \hat{p} \hat{U}^\dagger = e^{-\alpha}\hat{p} + m \dot{\alpha} e^{\alpha}\hat{q}. \label{9}
\end{array}\right.
\ee
This, in fact, corresponds to the classical canonical transformation given in Eq. (\ref{1}). From straightforward
algebra, the operator $\hat{U}$ satisfying Eqs. (\ref{8}) and (\ref{9}) is obtained in the form
\be
\hat{U} = \exp \left[{-\f{i}{2\hbar} m \dot{\alpha}e^{2\alpha} \hat{q}^2 }\right]
\exp \left[{\f{i}{2\hbar} \alpha (\hat{q}\hat{p} +\hat{p}\hat{q}) }\right]
\exp \left[{-\f{i}{\hbar}\beta \hat{p} }\right].   \label{10}
\ee
By inserting Eq. (\ref{10}) into Eq. (\ref{7}), the relation of $n$-th quantum state between both systems in $x-$space is calculated as
\be
\Psi_n (x,t) = \hat{U} \psi_n (x,t) = e^{\alpha/2} \exp {\left[-\f{i}{2\hbar}m \dot{\alpha} e^{2\alpha }x^2\right]}
 \psi_n (e^\alpha x - \beta,t).   \label{11}
\ee
This is a Schr\"{o}dinger solution of Eq. (\ref{6})  but not the
eigenfunction of the Hamiltonian $\hat{H}_2$.  Equation (\ref{11}) tells us that $n$-th quantum state of a
system is transformed to that of only another system by $\hat{U}$.
That is, Eq. (\ref{11}) is an adiabatic
change of the quantum state of the $\hat{H}_1$ system.

If an operator $\hat{I} (\hat{q},\hat{p},t)$ satisfies the Liouville-von Neumann equation,
\be
\f{d\hat{I}}{dt} = \f{\p \hat{I}}{\p t} + \f{1}{i \hbar} [\hat{I}, \hat{H}_2] = 0,   \label{12}
\ee
$\hat{I} (\hat{q},\hat{p},t)$ is then an invariant operator of the system \cite{Lewis}.
Generally, there are innumerable invariant operators satisfying Eq. (\ref{12})
in a system. Substituting Eq. (\ref{6}) into Eq. (\ref{12}), the invariant operator
quadratic in both $\hat{q}$ and $\hat{p}$ is obtained as
\be
\hat{I} (\hat{q},\hat{p},t) = \f{e^{-2\alpha} \hat{p}^2}{2m} + \f{m \dot{\alpha}^2 e^{2\alpha} \hat{q}^2}{2}
+ \f{\dot{\alpha}}{2}(\hat{q}\hat{p} +\hat{p}\hat{q}) + V(e^{\alpha}\hat{q}-\beta, R_0).   \label{13}
\ee
From Eq. (\ref{12}), we see that $\hat{I} (\hat{q},\hat{p},t)$ does not commute with $\hat{H}_2$.
This implies that the eigenstates of $\hat{I} (\hat{q},\hat{p},t)$ are different from those
of $\hat{H}_2$ for the time-dependent Hamiltonian system.
We also see that Eq. (\ref{13}) is obtained from the unitary transformation
of $\hat{H}_1$, i.e.,
\be
\hat{I} (\hat{q},\hat{p},t) = \hat{U}\hat{H}_1\hat{U}^\dagger  . \label{14}
\ee
Let us denote the eigenvalues and eigenfunctions of $\hat{I} (\hat{q},\hat{p},t)$ as $\lambda_n$ and
$|\phi_n (t) \rangle$,  respectively: \be
\hat{I} (\hat{q},\hat{p},t) |\phi_n (t) \rangle = \hat{U}\hat{H}_1\hat{U}^\dagger |\phi_n (t) \rangle
= \lambda_n |\phi_n (t) \rangle .  \label{15}
\ee
If we consider that $\hat{I} (\hat{q},\hat{p},t)$ as Hermitian, $\lambda_n$ is a real constant.
Since $\hat{U}$ is a unitary operator and $\hat{H}_1$ does not depend on time,
the eigenvalues and eigenfunctions of $\hat{H}_1$ should satisfy
\be
\hat{H}_1 \hat{U}^\dagger |\phi_n (t) \rangle = \lambda_n \hat{U}^\dagger |\phi_n (t) \rangle
= \lambda_n |\psi_n (t) \rangle .   \label{16}
\ee
From Eqs. (\ref{5}), (\ref{7}) and (\ref{16}), we see that
\be
|\Psi_n (t) \rangle = \hat{U} |\psi_n (t) \rangle = e^{-\f{i}{\hbar}\lambda_n t } |\phi_n (t) \rangle .
   \label{17}
\ee
In $x$-space, Eq. (\ref{17}) can be rewritten as
\ba
\Psi_n (x,t) &=& e^{-\f{i}{\hbar}\lambda_n t } \phi_n (x,t) = \hat{U} \psi_n (x,t) \nonumber \\
&=&\exp {\left[-\f{i}{2\hbar}m \dot{\alpha} e^{2\alpha }x^2\right]}
e^{\alpha/2} \psi_n (e^\alpha x - \beta,t) .  \label{18}
\ea
The Schr\"{o}dinger solution $\Psi_n (x,t)$ of the $\hat{H}_2$ system is not an eigenfunction of
$\hat{H}_2$, but rather an eigenfunction of the invariant operator $\hat{I} (\hat{q},\hat{p},t)$ of $\hat{H}_2$ system. 

\section{Extended adiabatic theorem}
  Berry's adiabatic theorem states that the quantum state $|\Psi_n (R(0),t) \rangle$ of the $\hat{H}_2 (R(0))$  system is
proportional to the $n$-th eigenket $|\Phi_{n(R(0))} (t) \rangle$ of the Hamiltonian $\hat{H}_2(R(0))$,
which is dependent on the external parameter $R(0)$. Specifically,
\be
|\Psi_n (R(0),t) \rangle = e^{- \f{i}{\hbar}\int_0^t E_n(t') dt' } e^{i\gamma_n(t)}
|\Phi_{n(R(0))} (t) \rangle ,  \label{19}
\ee
where $E_n(t)$ and $|\Phi_n (t) \rangle$ are eigenvalues and eigenkets of the Hamiltonian $\hat{H}_2$, respectively,
\be
\hat{H}_2 |\Phi_n (t) \rangle = E_n(t)|\Phi_n (t) \rangle ,  \label{20}
\ee
and Berry's phase $\gamma_n(t)$ is determined by
\be
\gamma_n(t) = i \int_{R(0)}^{R(t)}   \langle \Phi_{n(R(t'))} (t')| \nabla_R \Phi_{n(R(t'))} (t')
 \rangle dt' .  \label{21}
\ee  Here, $|\Psi_n (R(0),t) \rangle$ is not the solution of the Schr\"{o}dinger equation (6). If we replace
$|\phi_n (t) \rangle$ by $|\Phi_n (t) \rangle$ from Eq. (\ref{17}), $|\Psi_n (t) \rangle$ can no longer be Schr\"{o}dinger
 solution because $\hat{H_2} $ is dependent on time.
Only in the case that the eigenstate $\Phi _n (t)$ of Hamiltonian is the same as the eigenstate of the invariant operator,
Eq. (19) is valid as a Schr\"{o}dinger  solution.

To better understand the above theory, let us give an example. We choose $\hat{H}_1$ as the Hamiltonian of
the quantum simple harmonic oscillator and $\hat{H}_2$ as that of the quantum driven harmonic oscillator:

\be
\hat{H}_1 = \f{\hat{p}^2}{2m} + \f{m}{2} \omega^2 \hat{x}^2 ,  \label{23}
\ee
and
\be
\hat{H}_2 = \f{\hat{p}^2}{2m} + \f{m}{2} \omega^2 \hat{x}^2 - \hat{x} R(t) .  \label{24}
\ee
Recall that Eq. (\ref{18}) enables us to obtain the Schr\"{o}dinger solution of $\hat{H}_2$ system, provided that
the Schr\"{o}dinger solution of $\hat{H}_1$ system is known.
Since we can easily identify the Schr\"{o}dinger solution of $\hat{H}_1$ system in this case, we have the Schr\"{o}dinger
solution of the driven harmonic oscillator as
\be
\Psi_n (x,t) = \f{x_0^{-1/2}}{\pi^{1/4}\sqrt{2^n n!}} e^{-\left(\f{{\rm Im}{a_p}(t)}{x_0} \right)^2} e^{-i(n+1/2)\omega t}
e^{-\f{1}{2x_0^2}[x-a_p(t)]^2} H_n \left(\f{x-a_p(t)}{x_0} \right) ,  \label{25}
\ee
where $x_0 = \sqrt{\hbar/(m\omega)}$ and
\be
a_p(t) = \f{i}{m\omega} \int^t ds R(s) e^{-i\omega (t-s)} .  \label{26}
\ee
We also obtain, straightforwardly, the eigenfunctions and eigenvalues of $\hat{H}_2$ in the form
\ba
\Phi_{n(R(t))} (x,t) &=& \f{x_0^{-1/2}}{\pi^{1/4}\sqrt{2^n n!}}
e^{-\f{1}{2x_0^2}\left[x-\f{R(t)}{m\omega^2}\right]^2} H_n \left(\f{x-R(t)/(m\omega^2)}{x_0}
\right) ,  \label{27} \\
E_n(t) &=& \f{\hbar}{2} \left( n+\f{1}{2} \right) - \f{R^2(t)}{2m\omega^2} .  \label{28}
\ea
From a rigorous evaluation using Eq. (\ref{27}),  we have
\be
| \nabla_R \Phi_{n(R(t))} (t) \rangle = | \Phi_{n(R(t))-1} (t) \rangle  . \label{29}
\ee
This implies that $\gamma_n(t) = 0$, and from this we know that Eq. (\ref{19}) holds only for $ \hat{H}_2 (R(0))$ system.

Equation  (24) is not proportional to Eq. (26), i.e., the function $\Psi_{n(R(t))} (x,t) $ in Berry's
adiabatic theorem is not the Schr\"{o}dinger solution of the quantum driven harmonic oscillator.
Berry's adiabatic theorem should be extended in regard to our research as follows:  ``The quantum state,
$|\Psi_{n} (t) \rangle $ of the $ \hat{H}_2 $ system is proportional to the $n$-th eigenfunction of the
invariant operator, $\hat{I}(\hat{q},\hat{p},t )$, of  $\hat{H}_2 $ system, but not to the eigenket
$|\Phi_{n(R(t)}, t) \rangle $ of Hamiltonian $\hat{H}_2 (t)$." This revised theorem includes Berry's
original adiabatic theorem,  because $\hat{H}_2 (R(0))$ itself is an invariant operator when $R(t)=R(0)$. 

\section{Conclusion}
Here, we have derived a unitary operator that corresponds to a classical canonical transformation
connecting an adiabatic open system, that is affected by external circumstances, to a conserved system.
Because the canonical transformation that connects a general conserved system with
its corresponding adiabatic open system can be regarded as
a general treatment of adiabatic change classically, the derivation of quantum states using a unitary
transformation corresponding to adiabatic change
is a general treatment of a quantum adiabatic open system.
Thus, the results derived here are the counterparts of the classical results.
If Eq. (\ref{1}), which is a canonical transformation, has another form,
the adiabatic condition cannot be satisfied classically.
In this case, the connection fulfilled by the unitary operator corresponding to the canonical
transformation does not always give the same quantum state,
but can be different from the former one. Thus, the adiabatic theorem related to the connection of quantum states does not hold.

Though it is impossible to understand classically the quantum systems that have no classical counterpart,
like spin systems, we can derive relevant quantum states for adiabatic quantum systems from the eigenstates
of a quantum invariant operator.  In Eq. (\ref{1}), any type of canonical conjugate for $q(t)$ is acceptable provided
it has the form $p(t)+g(q,t)$.
In other words, there are innumerable Hamiltonians that give the same classical solution,
and each Hamiltonian has its particular canonical momentum.
The reason is that the canonical momenta cannot be distinguished
from the classical equations of motion,
whereas we can distinguish different positions. The fact that there exist many different canonical
momenta corresponding to a single classical canonical position tells us that there are numerous quantum
mechanical systems corresponding to a classical solution.
In other words, the canonical momenta are distinguishable from a quantum mechanical viewpoint.
The development of the theory presented here is possible for any type of canonical momentum.

\end{document}